# $\beta$-Ga$_2$O$_3$ on Insulator Field-effect Transistors with Drain Currents Exceeding 1.5 A/mm and Their Self-heating Effect

Hong Zhou, Kerry Maize, Gang Qiu, Ali Shakouri, Peide D. Ye [a]

*School of Electrical and Computer Engineering and Birck Nanotechnology Center,*

*Purdue University, West Lafayette, IN 47907, U.S.A.*

We have demonstrated that depletion/enhancement-mode $\beta$-Ga$_2$O$_3$ on insulator field-effect transistors can achieve a record high drain current density of 1.5/1.0 A/mm by utilizing a highly doped $\beta$-Ga$_2$O$_3$ nano-membrane as the channel. GOOI FET shows a high on/off ratio of $10^{10}$ and low subthreshold slope of 150 mV/dec even with 300 nm thick SiO$_2$. The enhancement-mode GOOI FET is achieved through surface depletion. An ultra-fast, high resolution thermo-reflectance imaging technique is applied to study the self-heating effect by directly measuring the local surface temperature. High drain current, low R$_c$, and wide bandgap make the $\beta$-Ga$_2$O$_3$ on insulator field-effect transistor a promising candidate for future power electronics applications.

[a]Author to whom correspondence should be addressed; electronic mail: yep@purdue.edu

Recently, β-Ga$_2$O$_3$ has shown its great promise for next generation high power device applications due to its ultra-wide bandgap of 4.6-4.9 eV.[1] This allows β-Ga$_2$O$_3$ to possess a corresponding empirical estimated electrical breakdown field (E$_{br}$) of 8 MV/cm, which is several times higher than GaN and SiC. Despite the very early development stage, depletion-mode (D-mode) β-Ga$_2$O$_3$ metal-oxide-semiconductor field-effect transistors (MOSFETs) have demonstrated a high blocking voltage of 750 V and an E$_{br}$ of 3.8 MV/cm [2,3] and enhancement-mode (E-mode) MOSFET has shown a breakdown voltage (BV) of more than 600 V.[4] β-Ga$_2$O$_3$ metal-semiconductor field-effect transistors (MESFETs) and Schottky barrier diodes (SBDs) have also demonstrated a BV of 257 V and 1000 V, respectively.[5,6] In addition to the excellent direct current (DC) breakdown characteristic, a high pulsed drain current (I$_D$) of 478 mA/mm has also been achieved for β-Ga$_2$O$_3$ MOSFETs.[7] Until very recently, radio frequency (RF) performance of β-Ga$_2$O$_3$ MOSFET was demonstrated, yielding cut off frequency and maximum oscillation frequency (f$_T$ and f$_{max}$) of 3.3 GHz and 12.9 GHz, respectively.[8] Meanwhile, β-Ga$_2$O$_3$ also has the advantage of a low-cost native bulk substrate that can be synthesized in large size through melt-grown Czochralski, edge defined film fed growth, and floating zone method.[9-14] However, a crucial disadvantage of this material is its low thermal conductivity of 0.1~0.3 W/cm·K depends on its various crystal orientation.[15,16] One of the approaches to solve the low thermal conductivity issue of β-Ga$_2$O$_3$ is to introduce a high thermal conductivity substrate rather than the β-Ga$_2$O$_3$ native substrate. Our previous study has demonstrated a high performance β-Ga$_2$O$_3$ on insulator field-effect transistors (GOOI FETs) by transferring β-Ga$_2$O$_3$ nano-membrane or nano-belts to SiO$_2$/Si substrate with SiO$_2$ thickness of 300 nm to mitigate gate-drain overlap breakdown and serve as gate dielectric as well.[17] The monoclinic structure of bulk β-Ga$_2$O$_3$ crystals would allow a facile cleavage into nano-membrane along the [100] direction even though β-Ga$_2$O$_3$ is not a van der Waals 2D material, possibly due to its large lattice constant of 12.23 Å along this [100] direction.[18]

To realize all merits of β-Ga$_2$O$_3$ as a power device, there are still some challenges to be encountered ahead. The reported D/E mode maximum drain current density (I$_{DMAX}$) for GOOI FETs are 600/450 mA/mm and continuous wave DC of 200 and 2 mA/mm for β-Ga$_2$O$_3$ homoepitaxial D/E mode MOSFETs, respectively.[19,20] How to further enhance the device performance in terms of higher I$_{DMAX}$ and lower on-resistance (R$_{on}$) and make them comparable to GaN and SiC technologies, remained to be demonstrated. In this letter, we have shown that by increasing the doping concentration of the β-Ga$_2$O$_3$ nano-membrane and further scale the device, the I$_{DMAX}$ of D/E-mode GOOI FETs can achieve a record value of 1.5/1.0 A/mm, which is around twice of previous record I$_{DMAX}$. Our results reveal that the contact resistance (R$_c$) can also be reduced by highly doped channel. Finally, we have evaluated the thermal effect of GOOI FETs and observed the pronounced self-heating effect by using an ultra-fast, high resolution thermo-reflectance (TR) imaging technique.

In our experiment, n-type Sn doping concentration (n) of β-Ga$_2$O$_3$ is determined to be 8.0×10$^{18}$ cm$^{-3}$ from Capacitance-Voltage (C-V) measurement.[21] The thickness of (100) β-Ga$_2$O$_3$ nano - membranes vary from 50 to 70 nm determined by atomic force microscopy (AFM) measurements with surface roughness of 0.33 nm, and the D/E-mode devices are achieved based on the thickness of the nano-membranes. More details about the device fabrication can be found in our previous work.[17] For comparison, devices with lower doping concentration of $3.0\times10^{18}$ cm$^{-3}$ were also fabricated. Figure 1 (a) and (b) are device schematic and AFM image of a fabricated device, respectively. The device electric characterizations were carried out with Keithley 4200 Semiconductor Parameter Analyzer. A Microsanj TR system with a high-speed LED pulse and a synchronized charge coupled device (CCD) camera was used for the thermal measurement.[22-25]

Figure 2 (a) presents the DC output characteristics ($I_D$-$V_{DS}$) of D-mode GOOI FETs. Devices have a channel length ($L_{CH}$, also source to drain spacing $L_{SD}$) of 0.3 μm, channel width (W) of 0.15 μm and channel thickness (t) of 70 nm for $8.0\times10^{18}$ cm$^{-3}$, and W = 0.6 μm and t = 100 nm for $3.0\times10^{18}$ cm$^{-3}$ device. The device dimensions are accurately determined by scanning electron microscopy (SEM). A record high $I_{DMAX}$ of 1.5 A/mm is obtained, which is more than 2 times of lower doping channel. Compared with our previous work with $L_{CH}$=0.87 μm and $3.0\times10^{18}$ cm$^{-3}$ doping concentration, the $I_{DMAX}$ of same doping device with $L_{CH}$=0.3 μm increased from 600 to 750 mA/mm for D-mode. D-mode GOOI FET with n of $3\times10^{18}$ cm$^{-3}$ and $L_{CH}$ of 0.3 μm has a $R_C$, $R_{SH}$ and $R_{on}$ of 4.5 Ω·mm, 5.6 kΩ/ and 11 Ω·mm, showing that the total contact resistance (2$R_C$) of 9 Ω·mm is dominate. Therefore, further scaling the $L_{CH}$ shows no significant effect in boosting the $I_{DMAX}$, which is mainly limited by the high $R_C$ of lower doped channel underneath the contacts. The much improved $I_{DMAX}$ of 1.5 A/mm mainly originates from the higher doping concentration induced lower $R_c$ rather than the simply channel length scaling.[26] The high doping device also has a much lower $R_{on}$ of 5.3 Ω·mm compared with that of 11 Ω·mm with $3.0\times10^{18}$ cm$^{-3}$ doping concentration and this significant lower $R_{on}$ is mostly from the much reduced $R_C$ of 1.7 Ω·mm. At low $V_{DS}$ regime, high doping $I_D$-$V_{DS}$ shows linear behavior with lower $R_c$ while the lower doping counterpart displays a Schottky-like contacts with higher $R_c$. Figure 2 (b) is the log-scale $I_D$-$g_m$-$V_{GS}$ transfer characteristics of the same D-mode GOOI FET with $8.0\times10^{18}$ cm$^{-3}$ doping concentration. This D-mode GOOI FET has a threshold voltage ($V_T$) of -135 V, extracted from the log-scale $I_D$-$V_{GS}$ at $V_{DS}$ = 1 V and $I_D$ = 0.1 mA/mm. A peak transconductance ($g_{max}$) of 9.2 mS/mm is achieved which is 2 times of the $g_{max}$ with lower doping concentration, showing the much improved $R_c$ of the higher doping concentration device. Figure 2 (c) and (d) depict the $I_D$-$V_{DS}$ output and $I_D$-$g_m$-$V_{GS}$ transfer characteristics of an E-mode GOOI FET with $L_{CH}$=0.3 µm, t = 55 nm and W=0.17 µm for $8.0\times10^{18}$ cm$^{-3}$, and t=75 nm and W=0.45 µm for $3.0\times10^{18}$ cm$^{-3}$ device also shown in figure 2 (c) as black dashed curves for comparison. Similar to D-mode devices, lower doped E-mode GOOI FETs have an increased $I_{dmax}$ from 450 mA/mm to 550 mA/mm when the $L_{CH}$ is scaled from 1.3 µm to 0.3 µm. Higher E-mode $I_{DMAX}$ is also achieved with higher doping concentration induced lower $R_c$ of 0.75 Ω·mm compared to that of 1.2 Ω·mm with lower doping concentration. A record high $I_{DMAX}$=1.0 A/mm for higher doping channel is obtained, which is more than 80% higher than lower doping channel. E-mode GOOI FET has a $V_T$ of 2 V determined from the $I_D$-$V_{GS}$ at $V_{DS}$=1 V and $I_D$=0.1 mA/mm. Unfortunately, no $I_D$ saturation is observed since applying higher $V_{DS}$ will lead to an abrupt $I_D$ increase and then device breakdown. The drain induced barrier lowering (DIBL) is extracted to be 0.73 and 0.38 V/V for D/E-modes devices, respectively. Finally, benefited from its wide-bandgap and high quality interface between β-Ga$_2$O$_3$ and SiO$_2$, both D/E-mode devices have achieved high on/off ratio of $10^{10}$ and low subthreshold slope (SS) of 150~165 mV/dec for 300 nm SiO$_2$.

The significant $V_T$ shift with respect to different β-Ga$_2$O$_3$ nano-membrane thickness is due to the surface depletion effect of the unpassivated GOOI FET surface. This is because Sn-doped β-Ga$_2$O$_3$ is a 3D semiconductor, which has dangling bonds and surface states on the device surface. Surface depletion could deplete the whole β-Ga$_2$O$_3$ nano-membrane tens of nanometers thick. That's the reason why E-mode GOOI FETs can also be realized in high doping β-Ga$_2$O$_3$ nano-membrane. The surface depletion effect is verified by using atomic layer deposition (ALD) to deposit 15 nm Al$_2$O$_3$ on top to passivate the top surface. As shown in figure 1, the $V_T$ is significantly shifted to the left for more than 70 V after the ALD passivation, showing the existence of top and bottom surface depletion on unpassivated GOOI FET surfaces. Similar surface or interface charges induced depletion effect was also observed by Moser et al.[7] Based on the surface depletion, we can obtain each surface depleted charge density ($n_s$) by using the TCAD C-V simulation to match the measured and simulated $V_T$ from E-mode devices with $V_T$ near zero. The $n_s$ is determined and simulated to be $1.2 \times 10^{13}$ cm$^{-2}$ and $2.2 \times 10^{13}$ cm$^{-2}$ for $3.0 \times 10^{18}$ cm$^{-3}$ and $8.0 \times 10^{18}$ cm$^{-8}$ nano-membranes with thickness of 80 nm and 55 nm, respectively. Therefore, the flat-band voltage ($V_{FB}$) for lower doped and high doped devices are determined to be 135 V and 235 V through the equation $V_{FB} = \Phi_{MS}/e - 2n_s/C_{ox}$, where $\Phi_{MS}$, e and $C_{ox}$ are gate-semiconductor work function difference, electron charge quantity and oxide capacitance, respectively. Higher $n_s$ for higher doping β-Ga$_2$O$_3$ nano-membrane is most likely related to the higher surface states with more Sn$^{4+}$ dopants. Therefore, the actual C-V curve and $I_D$-$V_{GS}$ curve are significantly shifted to the right compared with the ideal case without considering surface depletion. Figure 2 shows the simulated C-V curve for E-mode GOOI FET and the $V_T$ from C-V simulation is in good agreement with the $V_T$ from $I_D$-$V_{GS}$ characterization. Figure 3 is the simulated band diagram of the E-mode GOOI FET at $V_{GS}$ = 0 V with lower doping and high doping channels by considering surface depletion effect. The top and bottom depletion regions pull up the conduction band of β-$Ga_2O_3$ with very few carriers left behind in the nano-membrane.

As a low thermal conductivity material and also its substrate, the heat dissipation is a big issue for β-$Ga_2O_3$ devices which needs to be seriously considered. We have used an ultra-fast TR set-up to study its thermal property. The system has a high-speed LED pulse illumination and a CCD camera to image temperature dependent reflectance change. Briefly, the source/drain Au pads are illuminated through an LED (λ=530 nm) and the change in reflectance under bias is calibrated with Au thermo-reflectance coefficient to translate into temperature change. Figure 4 (a) is the CCD camera and TR image merged view of another D-mode GOOI FET device with $V_{DS}$=4 V and $V_{GS}$=0 V with $I_D$=0.3 A/mm. Figure 4 (b) shows the 3D view of the TR image along the channel length and channel width directions. Even at a low bias power regime (P=$V_{DS}$×$I_D$=1.2 W/mm), the local temperature has increased by 35 °C compared to room temperature or unbiased devices. It seems that there is a 'cold' channel between source and drain contact. This is because the thermal reflectance coefficient of β-$Ga_2O_3$ channel is more than 10 times lower than that of Au electrodes. The temperature measurement is only calibrated with Au surface and we use Au electrodes as the thermometer of the device. Higher power bias will increase device temperature more significantly, and lead to degrade electron mobility and reliability, and eventually breakdown the device. Other self-heating effects induced $I_D$ reduction was observed by Moser and Wong et al.[27,28] The work by applying large thermal conductivity substrates and advanced device structures to mitigate the self-heating effect for β-Ga$_2$O$_3$ FETs is on-going and will be reported elsewhere.

We have demonstrated record high I$_{DMAX}$ of 1.5/1.0 A/mm for D/E-mode GOOI FETs by increasing the β-Ga$_2$O$_3$ doping concentration from $3.0\times10^{18}$ to $8.0\times10^{18}$ cm$^{-3}$ and further scaling the channel length. The significant V$_T$ shift with respect to different β-Ga$_2$O$_3$ nano-membrane thickness is due to the surface depletion effect of the unpassivated GOOI FET surface. High on/off ratio of $10^{10}$ and low SS of 150 mV/dec are achieved. Self-heating effect is also directly observed with the TR measurement. GOOI FETs with wide bandgap, high I$_{DMAX}$ and low R$_c$ offer the promise in the power device applications if the low thermal conductivity issue can be solved.

The authors thank the technical guidance from the Sensors Directorate of Air Force Research Laboratory.

**Figure Captions**

Figure 1 (a) Schematic cross-section view of a GOOI FET with a 300 nm $SiO_2$ layer on Si substrate and (b) AFM image of fabricated GOOI FET with β-$Ga_2O_3$ nano-membrane thickness of 70 nm.

Figure 2 (a) and (c) are $I_D$-$V_{DS}$ output characteristics of D-mode and E-mode GOOI FETs with $3.0\times10^{18}$ and $8.0\times10^{18}$ cm$^{-3}$ doping channel, respectively. (b) and (d) are $I_D$-$g_m$-$V_{GS}$ transfer characteristics of D-mode and E-mode GOOI FETs with $8.0\times10^{18}$ cm$^{-3}$ doping channel, respectively. Record high $I_{DMAX}$ of 1.5 and 1.0 A/mm are demonstrated for D/E mode devices. Both D and E-mode devices have high on/off ratio of $10^{10}$ and low SS of 150~165 mV/dec for 300 nm $SiO_2$.

Figure 3 (a) $I_D$-$V_{GS}$ comparison between GOOI FETs with and without ALD passivation for β-$Ga_2O_3$ nano-membrane with doping concentration of $3.0\times10^{18}$ cm$^{-3}$, (b) Simulated C-V curve for E-mode GOOI FET at a β-$Ga_2O_3$ nano-membrane thickness of 80 nm and doping concentration of $3.0\times10^{18}$ cm$^{-3}$ after considering the top and bottom negative surface charge ($n_s$=$1.2\times10^{13}$ cm$^{-2}$) depletion effect. Band diagram and electron density distribution of E-mode GOOI FETs with surface negative charge depletion on (c) lower doping ($n_s$=$1.2\times10^{13}$ cm$^{-2}$) and (d) high doping ($n_s$=$2.2\times10^{13}$ cm$^{-2}$) β-$Ga_2O_3$ nano-membrane channels at $V_{GS}$=0 V.

Figure 4 (a) CCD camera and TR image merged temperature change on the source and drain sides of back-gate GOOI FETs. (b) 3D view of the TR image of GOOI FET. The device is biased at $V_{DS}$=4 V, $V_{GS}$=0 V, and $I_D$=0.3 A/mm. The temperature is calibrated for the S and D gold electrodes.

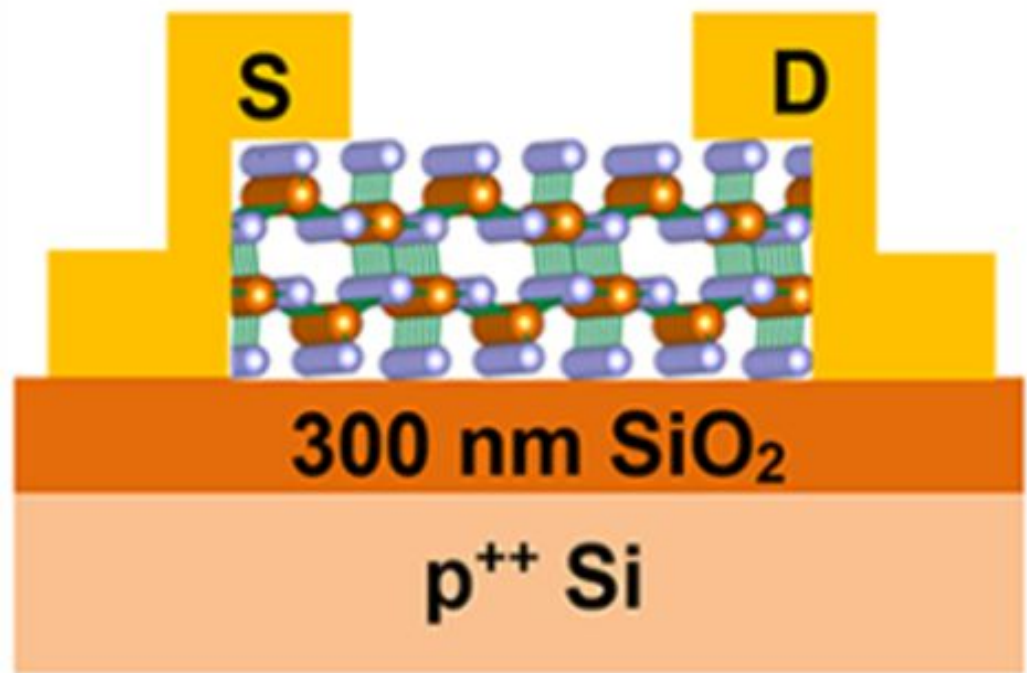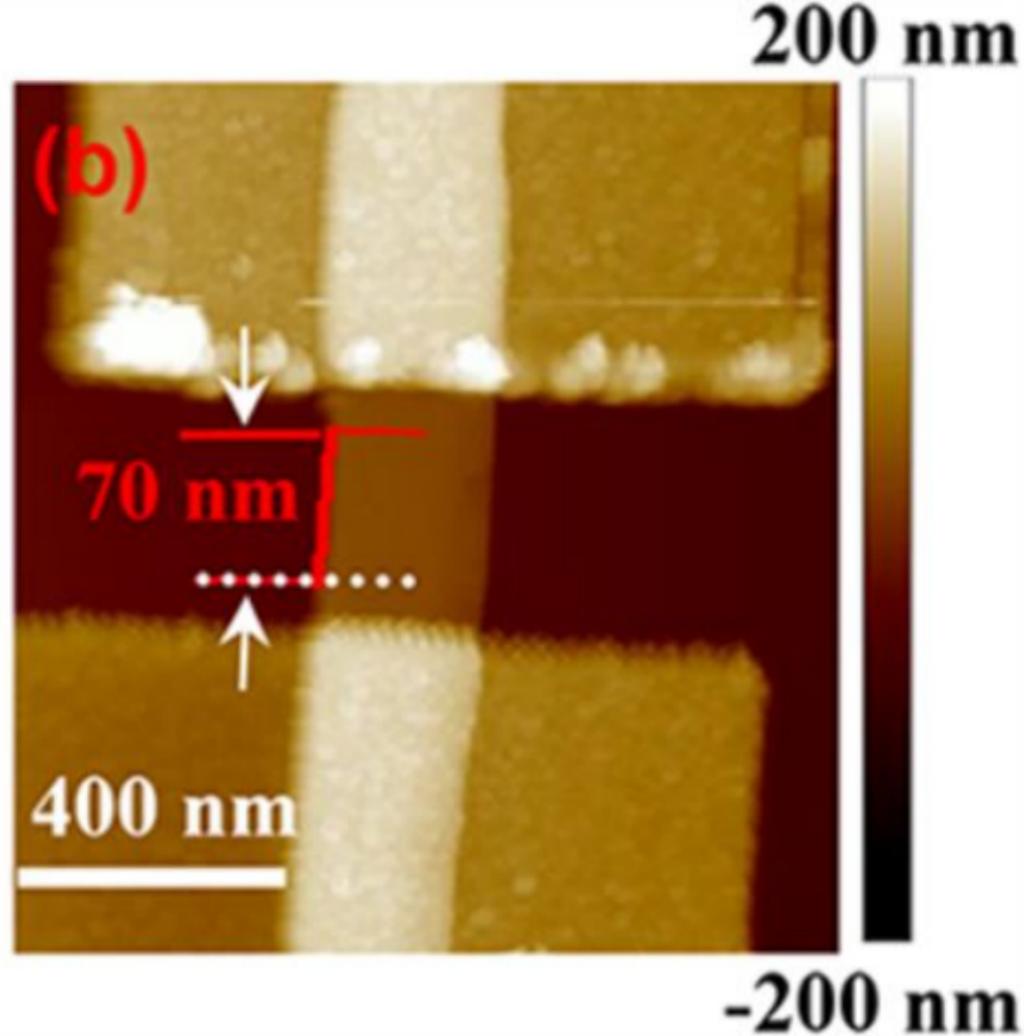

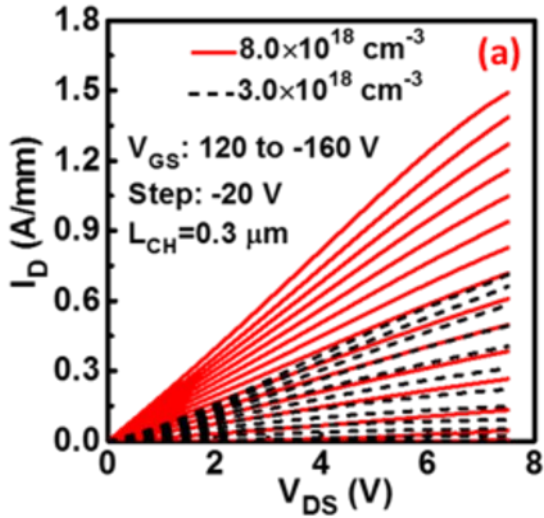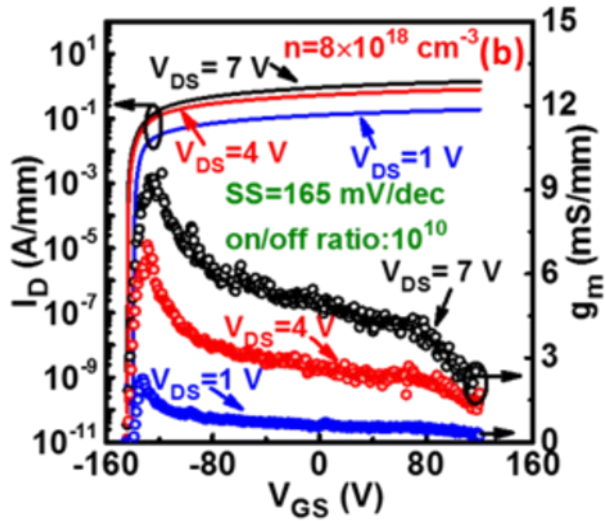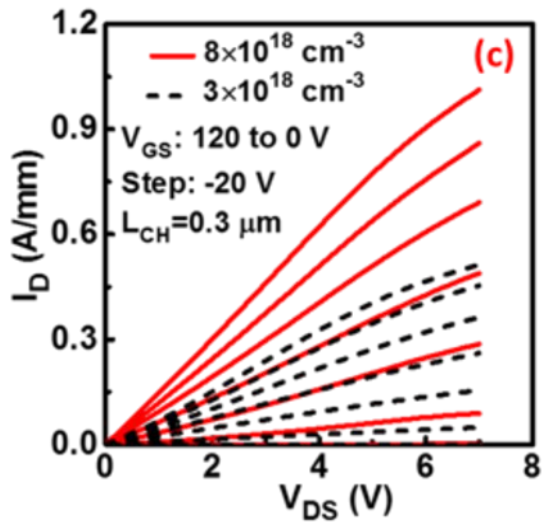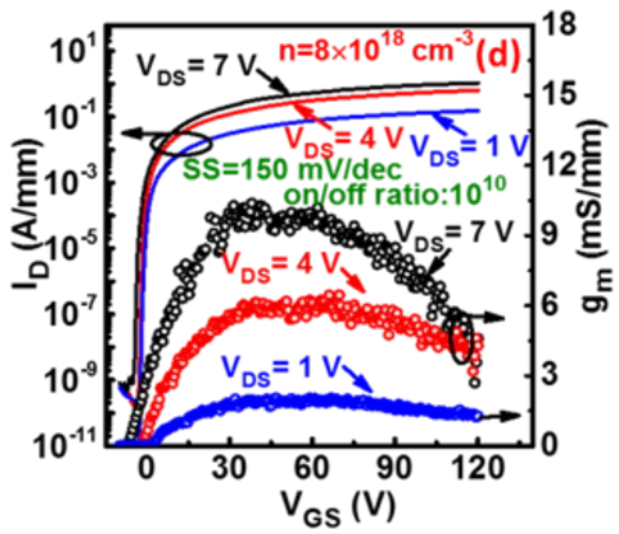

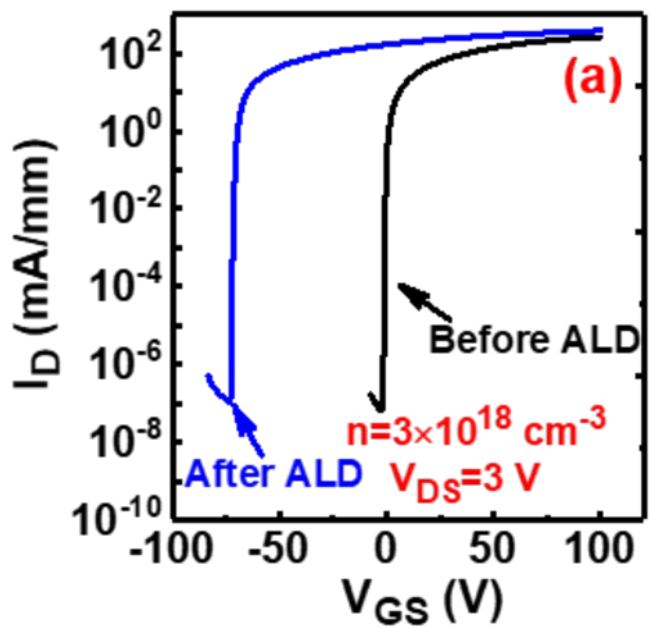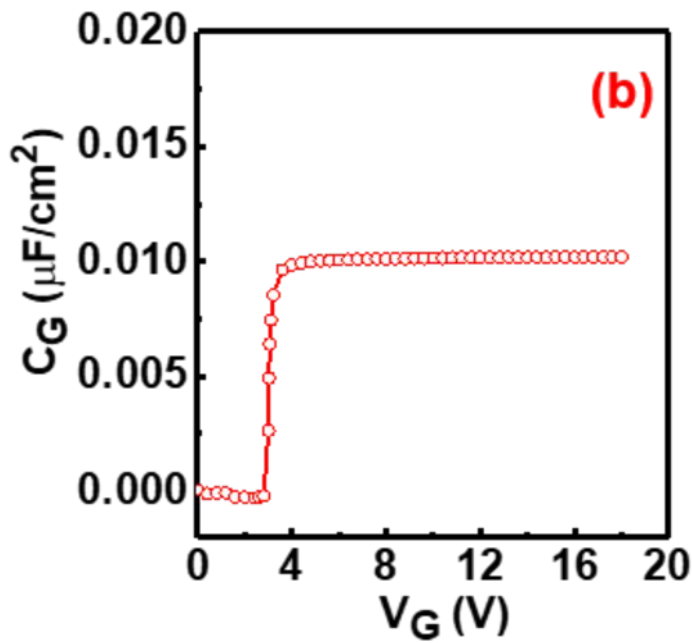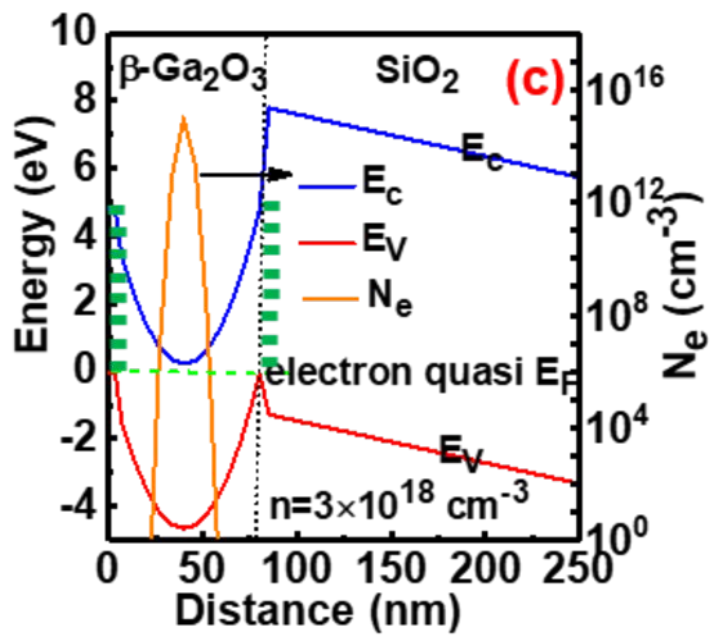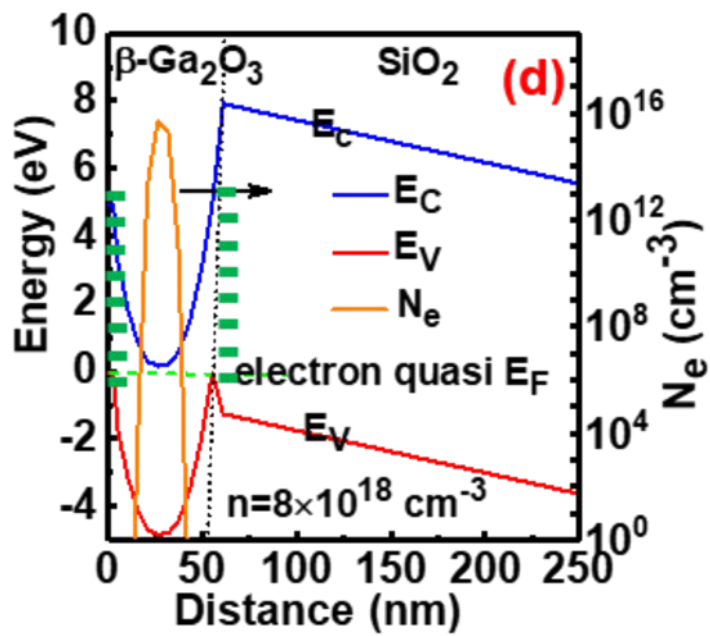

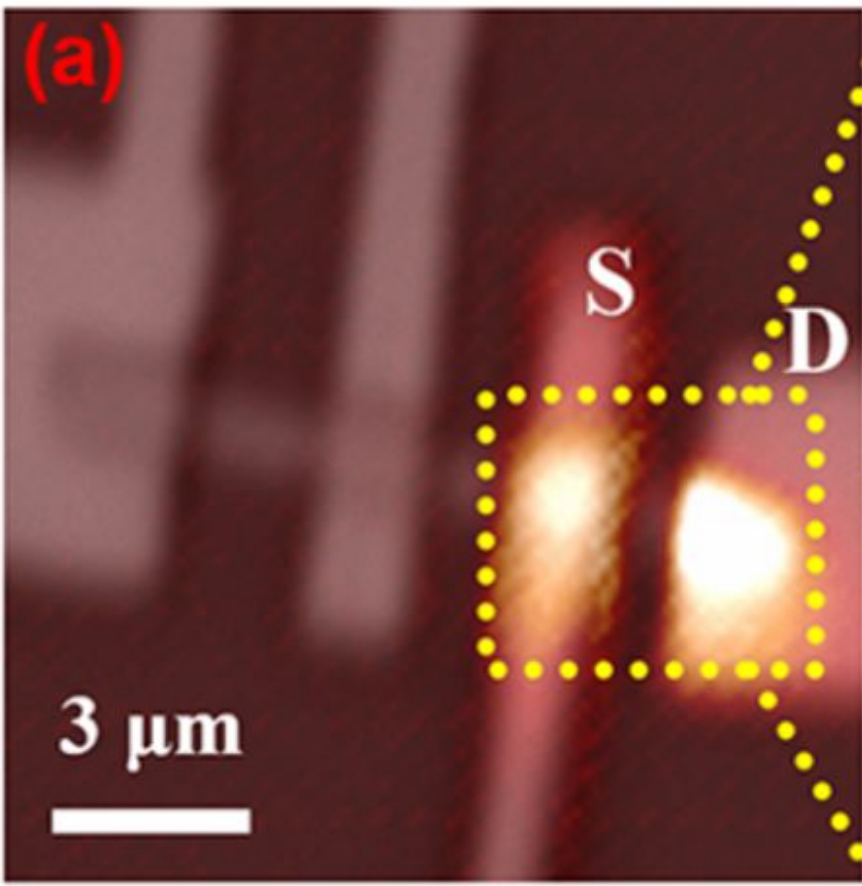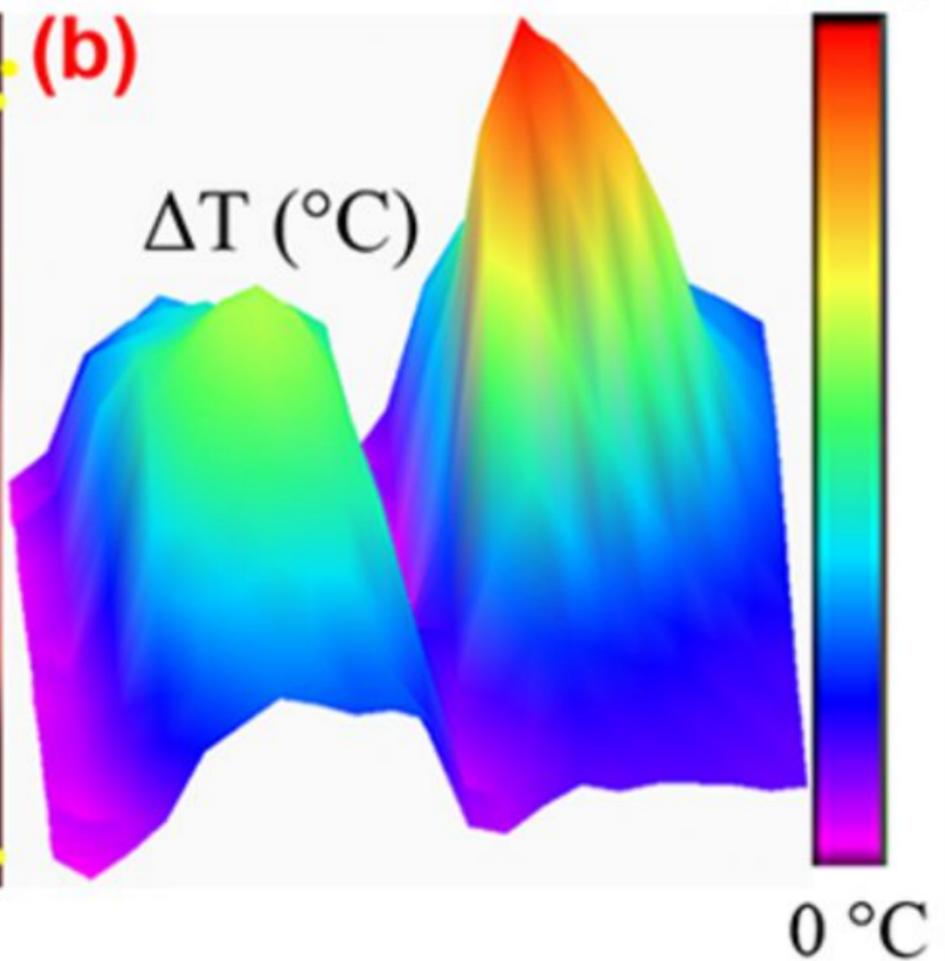